\documentclass[aps,pra,reprint,twocolumn,superscriptaddress,floatfix,nofootinbib,longbibliography]{revtex4-1}
\usepackage{graphicx,amsmath,amsfonts,amssymb,amsthm,xr}
\usepackage{epsfig,amsmath,amssymb,color,dsfont,upgreek,physics}
\usepackage{mathrsfs}
\usepackage{mathtools}
\usepackage{bbold}

\usepackage[bookmarks=true,colorlinks,linkcolor=OrangeRed,urlcolor=NavyBlue,citecolor=RoyalBlue]{hyperref}
\usepackage[dvipsnames]{xcolor}

\definecolor{mygold}{rgb}{0.93,0.69,0.13}

\definecolor{mypurple}{rgb}{0.49,0.18,0.56}

\definecolor{mygreen}{rgb}{0,0.5,0}

\definecolor{mygreen}{rgb}{0,0.5,0}
\definecolor{myred}{rgb}{0.7,0,0}

\begin{document}
\title{Dynamical quantum phase transitions in spin-$S$ $\mathrm{U}(1)$ quantum link models}
\author{Maarten Van Damme}
\affiliation{Department of Physics and Astronomy, University of Ghent, Krijgslaan 281, 9000 Gent, Belgium}
\author{Torsten V.~Zache}
\affiliation{Center for Quantum Physics, University of Innsbruck, 6020 Innsbruck, Austria}
\affiliation{Institute for Quantum Optics and Quantum Information of the Austrian Academy of Sciences, 6020 Innsbruck, Austria}
\author{Debasish Banerjee}
\affiliation{Theory Division, Saha Institute of Nuclear Physics, 1/AF Bidhan Nagar, Kolkata 700064, India}
\affiliation{Homi Bhabha National Institute, Training School Complex, Anushaktinagar, Mumbai 400094,India}
\author{Philipp Hauke} 
\affiliation{INO-CNR BEC Center and Department of Physics, University of Trento, Via Sommarive 14, I-38123 Trento, Italy}
\author{Jad C.~Halimeh}
\email{jad.halimeh@physik.lmu.de}
\affiliation{Department of Physics and Arnold Sommerfeld Center for Theoretical Physics (ASC), Ludwig-Maximilians-Universit\"at M\"unchen, Theresienstra\ss e 37, D-80333 M\"unchen, Germany}
\affiliation{Munich Center for Quantum Science and Technology (MCQST), Schellingstra\ss e 4, D-80799 M\"unchen, Germany}

\begin{abstract}
Dynamical quantum phase transitions (DQPTs) are a powerful concept of probing far-from-equilibrium criticality in quantum many-body systems. With the strong ongoing experimental drive to quantum-simulate lattice gauge theories, it becomes important to investigate DQPTs in these models in order to better understand their far-from-equilibrium properties. In this work, we use infinite matrix product state techniques to study DQPTs in spin-$S$ $\mathrm{U}(1)$ quantum link models. Although we are able to reproduce literature results directly connecting DQPTs to a sign change in the dynamical order parameter in the case of $S=1/2$ for quenches starting in a vacuum initial state, we find that for different quench protocols or different values of the link spin length $S>1/2$ this direct connection is no longer present. In particular, we find that there is an abundance of different types of DQPTs not directly associated with any sign change of the order parameter. Our findings indicate that DQPTs are fundamentally different between the Wilson--Kogut--Susskind limit and its representation through the quantum link formalism.
\end{abstract}
\date{\today}
\maketitle
\tableofcontents

\section{Introduction}
The field of far-from-equilibrium quantum many-body physics currently finds itself in a remarkable era of active quantum-simulation efforts seeking to realize evermore exotic phenomena with no true counterpart in equilibrium \cite{Hauke2012,Lamacraft2012,Georgescu_review,Eisert2015,Altman_review}. Naturally, such efforts align with the ultimate quest for possible dynamical quantum universality classes, for which various concepts of dynamical phase transitions have been proposed \cite{Berges2008,Heyl_review,Mori2018,Zvyagin2016,Marino_review}.

A prominent type of these, most frequently referred to as \textit{dynamical quantum phase transitions} (DQPTs), constitute an intuitive connection to thermal phase transitions \cite{Silva2008,Heyl2013,Heyl2014,Heyl2015}, and have been investigated in several quantum simulation platforms over the past few years \cite{Jurcevic2017,Flaeschner2018,Guo2019,Yang2019dqpt,Yang2019,Tian2020}. The essence of DQPTs lies in viewing the return probability amplitude---overlap of the time-evolved wave function with its initial state---as a boundary partition function, with complexified evolution time standing for the inverse temperature \cite{Heyl2013}. The negative logarithm of this quantity normalized by volume, called the return rate, is then the dynamical analog of the thermal free energy. Similarly to a thermal phase transition that occurs at a critical temperature where the thermal free energy (or any derivative thereof) exhibits a nonanalyticity, a DQPT arises at a \textit{critical time} at which the return rate is nonanalytic (see Sec.~\ref{sec:DQPT} for more details, and Ref.~\cite{Heyl_review} for an extensive review).

DQPTs have been demonstrated as a useful tool to extract far-from-equilibrium critical exponents \cite{Trapin2018,Halimeh2019b,Wu2019,Wu2020a,Wu2020b,Trapin2020,Halimeh2020}, and as a probe of the quasiparticles of target models \cite{Jafari2019,Halimeh2018a,Defenu2019}. Though initially discovered in one-dimensional integrable free-fermionic models \cite{Heyl2013,Budich2016,Dutta2017}, they have since been shown to be ubiquitous in many-body systems, arising in nonintegrable \cite{Karrasch2013,Vajna2014,Andraschko2014,Halimeh2017,Zunkovic2018}, higher-dimensional \cite{Weidinger2017,Heyl2018,DeNicola2019,Hashizume2018,Hashizume2020}, and mean-field models \cite{Homrighausen2017,Lang2017,Lang2018}. 

Recently, DQPTs have also been studied in gauge theories \cite{Zache2019,Huang2019}, a class of quantum many-body models describing the interactions between dynamical matter and gauge fields through local constraints enforced by the underlying gauge symmetries \cite{Weinberg_book,Gattringer_book,Zee_book}. Lattice gauge theories (LGTs) have recently been at the center of a number of impressive quantum-simulation experiments \cite{Martinez2016,Muschik2017,Bernien2017,Klco2018,Kokail2019,Goerg2019,Schweizer2019,Mil2020,Klco2020,Yang2020,Zhou2021}, and there is great interest in advancing these setups to quantum-simulate more complex gauge theories \cite{Wiese_review,Pasquans_review,Alexeev_review,aidelsburger2021cold,zohar2021quantum,klco2021standard,homeier2020mathbbz2}. Though initially a tool to address nonperturbative regimes in high-energy physics \cite{Gattringer_book}, LGTs have proven to be formidable venues for the realization of exotic far-from-equilibrium phenomena pertinent to condensed matter physics. Prominent examples include the ergodicity-breaking paradigms of disorder-free localization \cite{Smith2017,Brenes2018} and quantum many-body scars \cite{Bernien2017,Turner2018,Surace2020,Banerjee2021} that carry deep connections to fundamental questions on the thermalization of isolated quantum systems and the eigenstate thermalization hypothesis \cite{Deutsch1991,Srednicki1994,Rigol2006,Rigol2007,Rigol_2008,Rigol_review,Deutsch_review}.

In this light, it is important to further investigate DQPTs in LGTs, and, in particular, variations thereof that are pertinent to realizations in quantum simulators. In this vein, the paradigmatic $(1+1)-$dimensional spin-$S$ $\mathrm{U}(1)$ quantum link model (QLM) stands out \cite{Mil2020,Yang2020,Zhou2021}. It is a lattice version of the Schwinger model from quantum electrodynamics where the gauge fields of infinite-dimensional Hilbert space in the latter are represented by spin-$S$ operators of a finite-dimensional Hilbert space that is amenable for experimental implementations \cite{Wiese_review,Chandrasekharan1997}. Indeed, large-scale experimental realizations of the spin-$1/2$ $\mathrm{U}(1)$ QLM have recently been implemented to directly observe gauge invariance~\cite{Yang2020} and thermalization dynamics~\cite{Zhou2021}. In Ref.~\cite{Huang2019}, it has been shown for quenches starting in a vacuum state of the spin-$1/2$ $\mathrm{U}(1)$ QLM and ending across the critical point that DQPTs exhibit a one-to-one connection to the order parameter changing sign. Such a connection was first found in the XXZ chain for quenches starting in the symmetry-broken phase and ending across the critical point~\cite{Heyl2014}. 

In this work, we numerically simulate the quench dynamics in spin-$S$ $\mathrm{U}(1)$ QLMs using the infinite matrix product state technique (iMPS) based on the time-dependent variational principle (TDVP) \cite{Haegeman2011,Haegeman2013,Haegeman2016,Vanderstraeten2019,MPSKit}, which works directly in the limit of infinite system size. This allows us to reliably detect DQPTs and classify their type by studying the corresponding matrix product state (MPS) transfer matrices \cite{Zauner2015,Zauner2017}. We find that the physics of DQPTs is richer than what was found in Ref.~\cite{Huang2019} by considering different quenches at $S=1/2$, and any nontrivial quench at larger $S$. We find three main types of DQPTs, and their connection to a change of sign in the order parameter is shown not to be present in general.

The rest of the paper is organized as follows: In Sec.~\ref{sec:DQPT}, we will give a brief overview of DQPTs, their calculation in iMPS, and a glossary involving the new terms we introduce in this work. In Sec.~\ref{sec:U1QLM}, we discuss the spin-$S$ $\mathrm{U}(1)$ QLM. We present our numerical results for different quench protocols and values of the link spin length $S$ in Sec.~\ref{sec:QuenchDynamics}. We finally conclude and provide an outlook in Sec.~\ref{sec:conc}.

\section{Dynamical quantum phase transitions}\label{sec:DQPT}
The seminal work of Heyl \textit{et al.}~\cite{Heyl2013} pointed out that the return probability amplitude $\braket{\psi_0}{\psi(t)}=\bra{\psi_0}e^{-i\hat{H}t}\ket{\psi_0}$ is a boundary partition function with complexified time $it$ representing inverse temperature. As such, the quantity $-2L^{-1}\ln\lvert\braket{\psi_0}{\psi(t)}\rvert$ can be considered as a dynamical analog of the thermal free energy when the system of size $L$ is prepared in the initial state $\ket{\psi_0}$ and subsequently quenched by the Hamiltonian $\hat{H}$. In the thermodynamic limit $L\to\infty$, this dynamical free energy can exhibit nonanalytic behavior at \textit{critical evolution times} $t$. Nonanalyticities in the return rate have been dubbed DQPTs \cite{Heyl_review,Mori2018,Zvyagin2016}.

A slightly modified formulation of the return rate can be defined when the initial state resides in a degenerate ground-state manifold \cite{Zunkovic2018}, as is the case in all the scenarios considered below in Sec.~\ref{sec:QuenchDynamics}. Assume we prepare our system in one of two doubly degenerate ground states in the $\mathbb{Z}_2$ symmetry-broken (ordered) phase of some initial Hamiltonian $\hat{H}_0$. Let us call these ground states $\ket{\psi_0^\pm}$, which correspond to a positive or negative order parameter, respectively, and suppose that the system is initialized in $\ket{\psi_0^+}$. Upon quenching with a final Hamiltonian $\hat{H}$, we can then write
\begin{subequations}
\begin{align}\label{eq:RR}
    r(t)&=\min\big\{\lambda^+_1(t),\lambda^-_1(t)\big\},\\\label{eq:primary_RR}
    \lambda^+_1(t)&=-\lim_{L\to\infty}\frac{1}{L}\ln\big\lvert\braket{\psi_0^+}{\psi(t)}\big\rvert^2,\\\label{eq:secondary_RR}
    \lambda^-_1(t)&=-\lim_{L\to\infty}\frac{1}{L}\ln\big\lvert\braket{\psi_0^-}{\psi(t)}\big\rvert^2,
\end{align}
\end{subequations}
where the total return rate~\eqref{eq:RR} is the minimum of the primary return rate~\eqref{eq:primary_RR} onto the initial state $\ket{\psi_0^+}$ and the secondary return rate~\eqref{eq:secondary_RR} onto the other degenerate ground state $\ket{\psi_0^-}$ of $\hat{H}_0$. The subscript of $\lambda^\pm_1(t)$ will become clear in the context of iMPS, which we elucidate in the following.

\subsection{Calculation in infinite matrix product states}\label{sec:DQPT_iMPS}
In iMPS, the physical quantum transfer matrix that one obtains from a path-integral formulation of the return probability amplitude $\braket{\psi_0^\pm}{\psi(t)}$ is approximated by the MPS transfer matrix $\mathcal{T}^\pm(t)$ \cite{Zauner2015}. One can then define the rate-function branches \cite{Zauner2017}
\begin{align}
    \lambda^\pm_n(t)=-2\ln\lvert\varepsilon^\pm_n(t)\rvert,
\end{align}
where $\varepsilon^\pm_n(t)$ are the eigenvalues of $\mathcal{T}^\pm(t)$ in descending order: $\lvert\varepsilon^\pm_1(t)\rvert\geq\lvert\varepsilon^\pm_2(t)\rvert\geq\ldots\geq\lvert\varepsilon^\pm_\mathcal{D}(t)\rvert$, where $\mathcal{D}$ is the MPS bond dimension. Thus, the (primary or secondary) return rate $\lambda^\pm_1(t)$ is determined by the dominant eigenvalue $\varepsilon^\pm_1(t)$ of $\mathcal{T}^\pm(t)$, and the following relation always holds: $\lambda^\pm_1(t)\leq\lambda^\pm_2(t)\leq\ldots\leq\lambda^\pm_\mathcal{D}(t)$.

Now we can see that different types of DQPTs can arise in our framework, which we will define in the following.

\subsection{Glossary}\label{sec:DQPT_glossary}
\begin{figure}[t!]
	\centering
	\includegraphics[width=.48\textwidth]{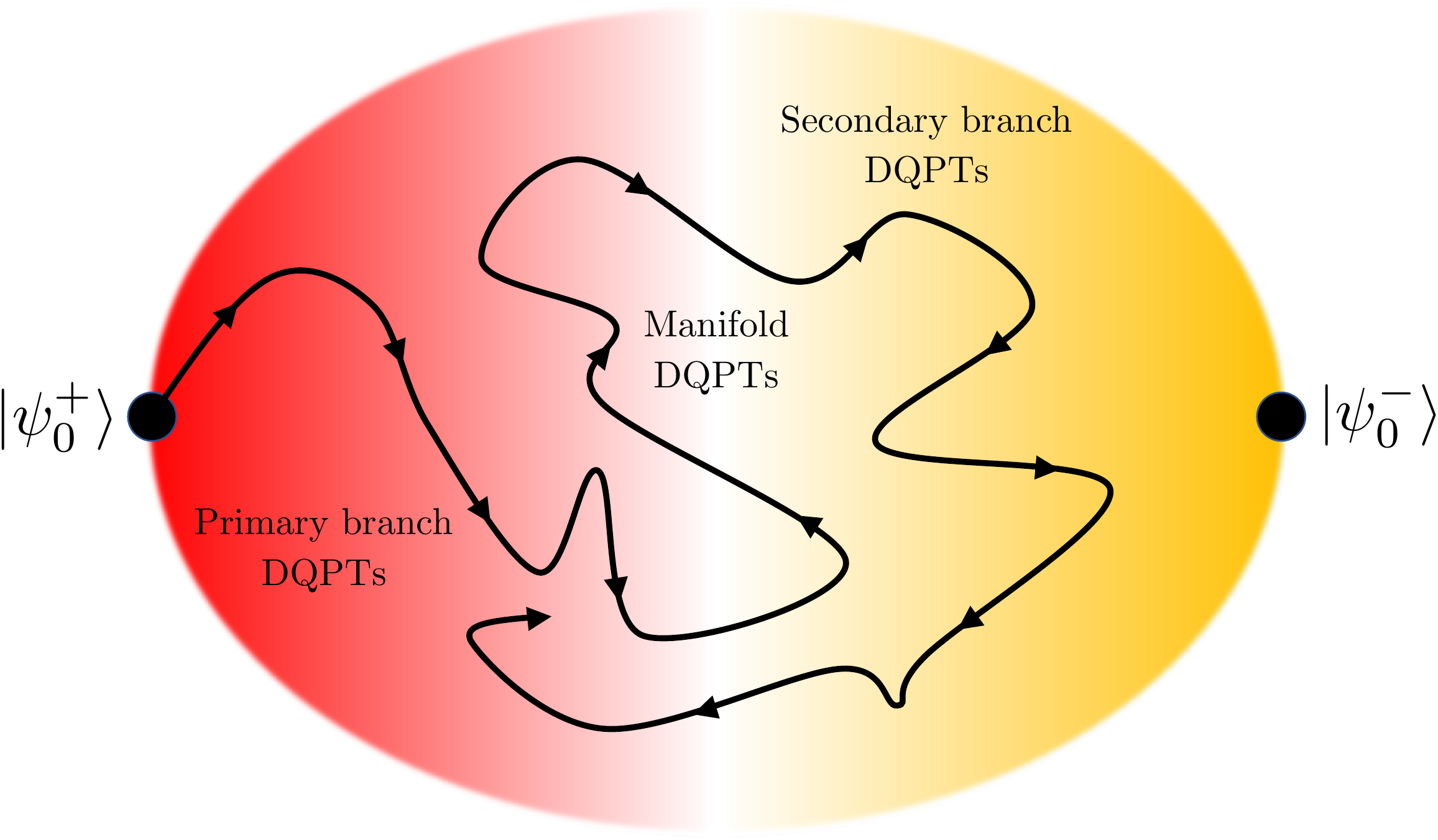}
	\caption{(Color online). Illustration of DQPT behavior in the quench dynamics of spin-$S$ $\mathrm{U}(1)$ QLMs. The $\mathbb{Z}_2$ symmetry-broken phase of the latter hosts two doubly degenerate ground states, $\ket{\psi_0^+}$ and $\ket{\psi_0^-}$. Starting in $\ket{\psi_0^+}$ and quenching across the equilibrium quantum critical point, rich dynamical criticality arises in the total return rate $r(t)$~\eqref{eq:RR} as the system explores the Hilbert space in the wake of the quench. When $r(t)$ switches from the primary return rate $\lambda_1^+(t)$~\eqref{eq:primary_RR} onto $\ket{\psi_0^+}$ to the secondary return rate $\lambda_1^-(t)$~\eqref{eq:secondary_RR} onto $\ket{\psi_0^-}$ or vice versa, a \textit{manifold DQPT} appears at a critical time $t_c$ when $r(t_c)=\lambda_1^+(t_c)=\lambda_1^-(t_c)$. Furthermore, DQPTs can also appear in the form of nonanalyticities in $r(t)$ that are not related to a switch between $\lambda_1^+(t)$ and $\lambda_1^-(t)$. Such a \textit{branch DQPT} occurs at critical times $t_c$ when $r(t_c)=\lambda_1^\alpha(t_c)=\lambda_2^\alpha(t_c)$, where $\lambda_1^\alpha(t_c)=\min\big\{\lambda_1^+(t_c),\lambda_1^-(t_c)\big\}$ and $\lambda_2^\alpha(t)$ is the second rate-function branch; see Sec.~\ref{sec:DQPT_iMPS} for details.}
	\label{fig:illustration} 
\end{figure}
\textit{\textbf{Manifold DQPTs}} are nonanalyticities appearing at a critical time $t_c$ in $r(t)$ due to intersections between $\lambda^+_1(t)$ and $\lambda^-_1(t)$, i.e., $r(t_c)=\lambda^+_1(t_c)=\lambda^-_1(t_c)$. These are not related to level crossings in the spectra of the MPS transfer matrices.

\textit{\textbf{Branch DQPTs}} are nonanalyticities appearing in $r(t)$ at critical times $t_c$ where the two lowest rate-function branches intersect:  $r(t_c)=\lambda^\alpha_1(t_c)=\lambda^\alpha_2(t_c)$, where $\lambda^\alpha_1(t_c)=\min\{\lambda^+_1(t_c),\lambda^-_1(t_c)\}$. These DQPTs are a direct result of level crossings in the spectrum of the MPS transfer matrix $\mathcal{T}^\alpha(t)$ between its two largest eigenvalues $\varepsilon_1^\alpha(t)$ and $\varepsilon_2^\alpha(t)$ at $t_c$. This manifests as an intersection between $\lambda^\alpha_1(t)$ and $\lambda^\alpha_2(t)$ at the critical time $t_c$. When $\alpha=+$, we shall refer to such a nonanalyticity in $r(t)$ as a \textit{primary} branch DQPT, while when $\alpha=-$, we shall refer to it as a \textit{secondary} branch DQPT, corresponding to the dominant component return rate in which the nonanalyticity occurs.

\section{Spin-$S$ $\mathrm{U}(1)$ quantum link model}\label{sec:U1QLM}
The $(1+1)-$D spin-$S$ $\mathrm{U}(1)$ QLM is described by the Hamiltonian \cite{Wiese_review,Chandrasekharan1997,Hauke2013,Yang2016,Kasper2017}
\begin{align}\nonumber
    \hat{H}=\sum_{j=1}^L\bigg[&\frac{J}{2a\sqrt{S(S+1)}}\big(\hat{\sigma}^-_j\hat{s}^+_{j,j+1}\hat{\sigma}^-_{j+1}+\text{H.c.}\big)\\\label{eq:H}
    &+\frac{\mu}{2}\hat{\sigma}^z_j+\frac{\kappa^2a}{2}\big(\hat{s}^z_{j,j+1}\big)^2\bigg].
\end{align}
This is a lattice version of $(1+1)-$D quantum electrodynamics (QED) where the gauge field is represented by the spin-$S$ operator $\hat{s}^\pm_{j,j+1}$ at the link between matter sites $j$ and $j+1$. Upon a Jordan--Wigner transformation, the fermionic matter creation and annihilation operators are represented by the Pauli matrices $\hat{\sigma}^\pm_j$ at matter site $j$. Additionally, we have employed a particle-hole transformation \cite{Hauke2013}. The matter occupation at site $j$ is given by the operator $\hat{n}_j=\big(\hat{\sigma}^z_j+\mathds{1}\big)/2$. The mass of the matter field is given by $\mu$, and the electric-field coupling strength is $\kappa$. The energy scale is set by $J=1$ throughout the paper. 

The principal property of Eq.~\eqref{eq:H} is the gauge symmetry generated by the discrete operator
\begin{align}
    \hat{G}_j=(-1)^j\big(\hat{n}_j+\hat{s}^z_{j-1,j}+\hat{s}^z_{j,j+1}\big),
\end{align}
which can be viewed as a discretized version of Gauss's law, imposing a local constraint on the electric-field configuration at the two links neighboring a matter site depending on the matter occupation at it. The gauge symmetry of Eq.~\eqref{eq:H} is encoded in the commutation relations $\big[\hat{H},\hat{G}_j\big]=0,\,\forall j$. Throughout this work, we will work in the ``physical" sector of Gauss's law: $\hat{G}_j\ket{\phi}=0,\,\forall j$.

Although a lattice version of QED with gauge fields represented by spin-$S$ operators, Eq.~\eqref{eq:H} has been shown to achieve the quantum field theory limit of QED at finite $a$ and relatively small $S$ both in \cite{Buyens2017,Banuls2018,Banuls2020,zache2021achieving} and out of equilibrium \cite{halimeh2021achieving}. Large-scale quantum simulations of Eq.~\eqref{eq:H} at $S=1/2$ have also recently been performed for a mass ramp \cite{Yang2020} and to probe thermalization dynamics in the wake of a global quench \cite{Zhou2021}.

\section{Quench dynamics}\label{sec:QuenchDynamics}
We will now present our main numerical results obtained from iMPS. For our most stringent calculations, we find within the considered evolution times that good convergence is achieved with a time-step of $0.001/J$ and a maximal bond dimension of $\mathcal{D}_\text{max}=550$. We are interested in two key quantities, the return rate defined in Eq.~\eqref{eq:RR} and the order parameter (electric flux)
\begin{align}\label{eq:flux}
    \mathcal{E}(t)=\frac{1}{L}\sum_{j=1}^L(-1)^{j+1}\bra{\psi(t)}\hat{s}^z_{j,j+1}\ket{\psi(t)},
\end{align}
where $\ket{\psi(t)}=e^{-i\hat{H}t}\ket{\psi_0^+}$.

\subsection{Quench from a $\mathbb{Z}_2$ symmetry-broken product state}\label{sec:InitProdState}
Let us first consider quenches starting in a $\mathbb{Z}_2$ symmetry-broken product state. In terms of the spin-$S$ $\mathrm{U}(1)$ QLM, this entails preparing the system in one of two doubly degenerate ground states of Eq.~\eqref{eq:H} at $\mu\to\infty$ for half-integer $S$ or $\mu\to-\infty$ for integer $S$, with real $\kappa\neq0$. In terms of the eigenvalues $n_j$ and $s^z_{j,j+1}$ of the matter occupation $\hat{n}_j$ and electric-field $\hat{s}^z_{j,j+1}$ operators, respectively, the two-site two-link unit cell representations of these product states are then  $\ket{n_j,s^z_{j,j+1},n_{j+1},s^z_{j+1,j+2}}=\ket{0,+1/2,0,-1/2}$ for half-integer $S$ and $\ket{n_j,s^z_{j,j+1},n_{j+1},s^z_{j+1,j+2}}=\ket{1,0,1,-1}$ for integer $S$. Let us denote the latter as $\ket{\psi_0^+}$, in which we prepare our system at $t\leq0$. The other degenerate product ground state $\ket{\psi_0^-}$ is then $\ket{n_j,s^z_{j,j+1},n_{j+1},s^z_{j+1,j+2}}=\ket{0,-1/2,0,+1/2}$ for half-integer $S$ and $\ket{n_j,s^z_{j,j+1},n_{j+1},s^z_{j+1,j+2}}=\ket{1,-1,1,0}$ for integer $S$. In the case of half-integer $S$, $\ket{\psi_0^\pm}$ represent the degenerate vacua of the $\mathrm{U}(1)$ QLM at nonzero $\kappa$, while for integer $S$ they are the degenerate charge-proliferated product states. 

At $t=0$, we quench the initial state $\ket{\psi_0^+}$ with $\hat{H}$ of Eq.~\eqref{eq:H} at $\mu=\pm0.655J/\big[6a\sqrt{S(S+1)}\big]$ (positive sign for half-integer $S$ and negative for integer $S$) and $\kappa=0.1\sqrt{J}$, which ensures that the quench is across the equilibrium quantum critical point for all considered values of $S$. Henceforth, we shall set $a=1$ in all our calculations.

\begin{figure}[t!]
	\centering
	\includegraphics[width=.48\textwidth]{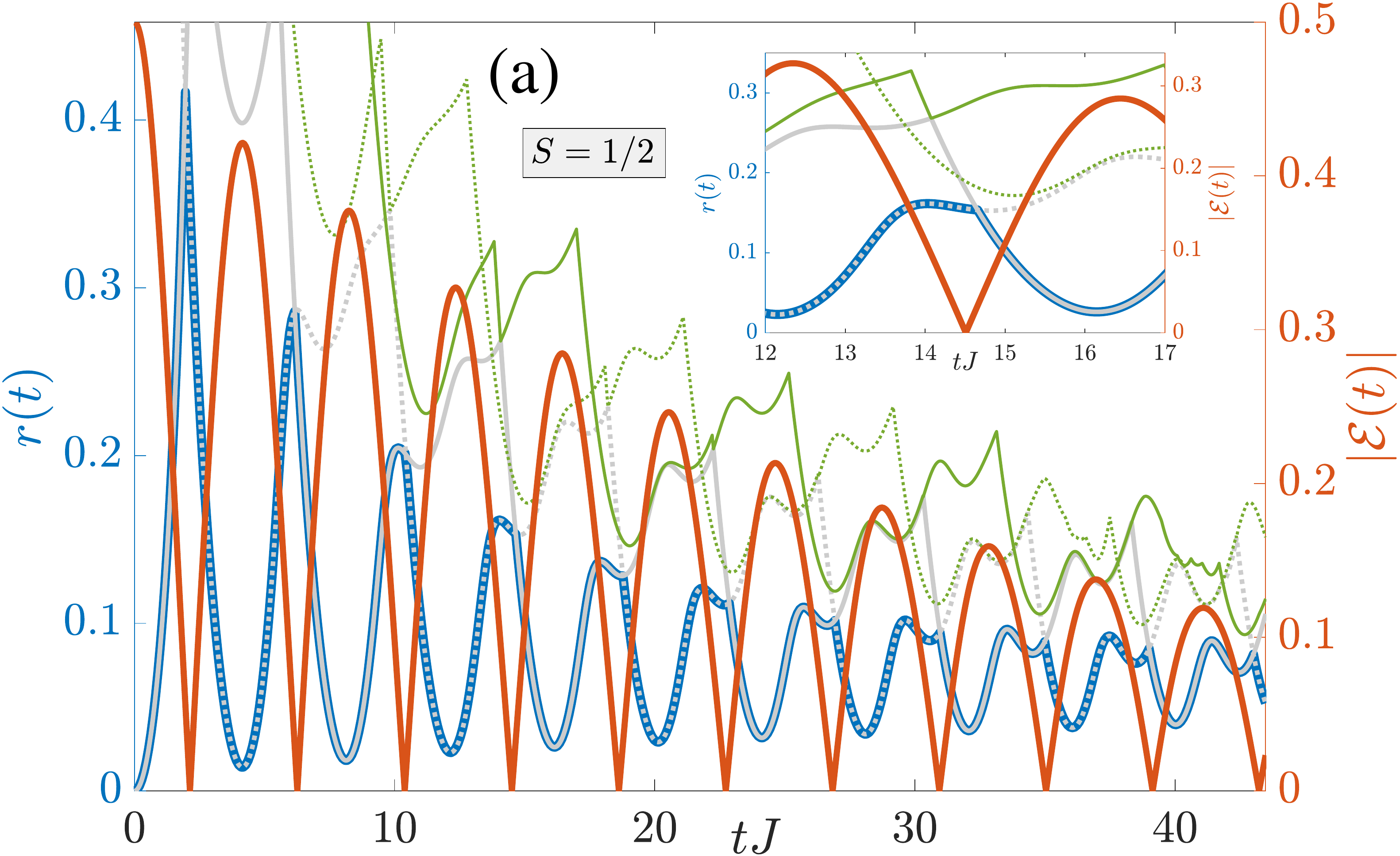}\\
	\vspace{1.1mm}
	\includegraphics[width=.48\textwidth]{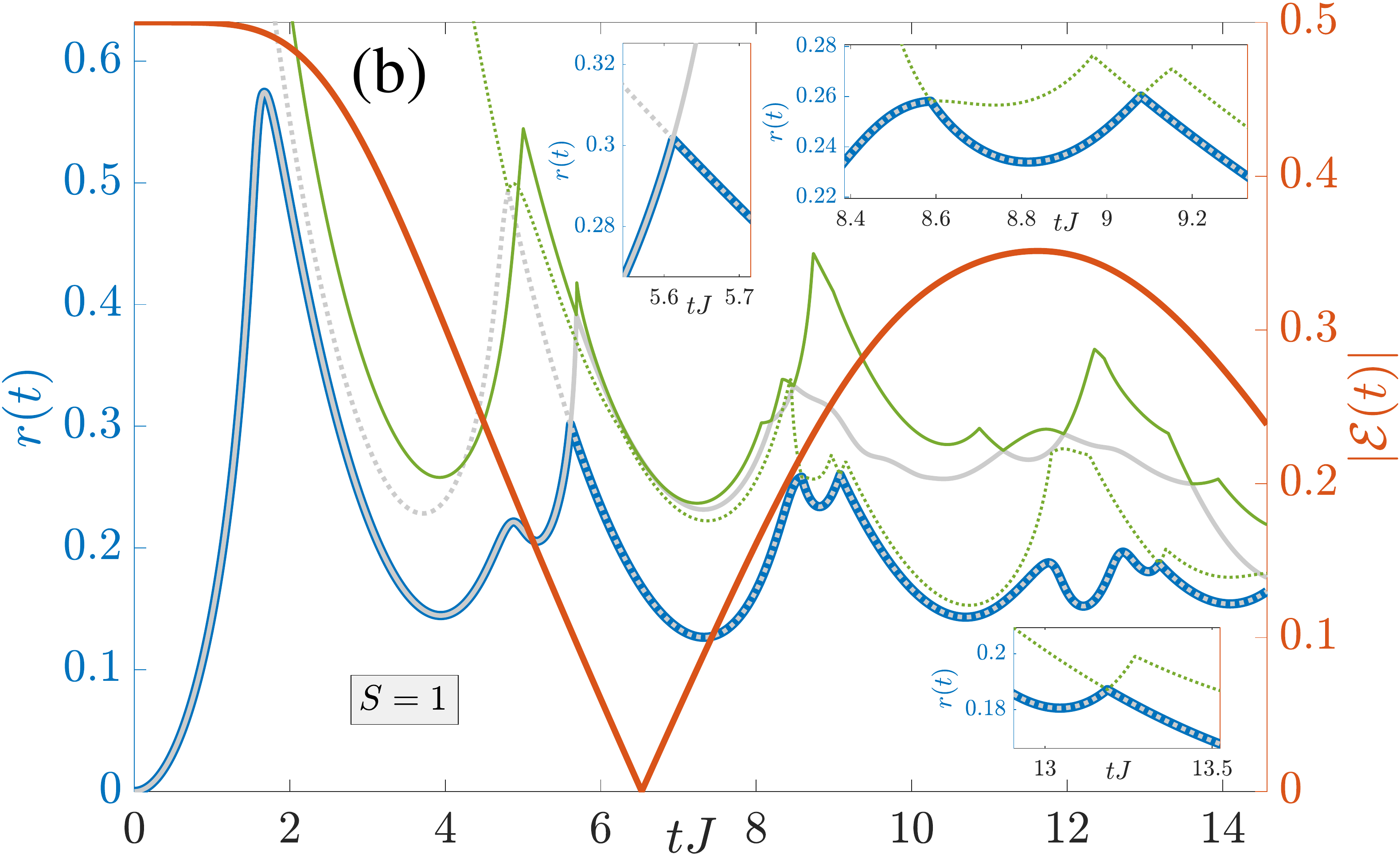}\\
	\vspace{1.1mm}
	\includegraphics[width=.48\textwidth]{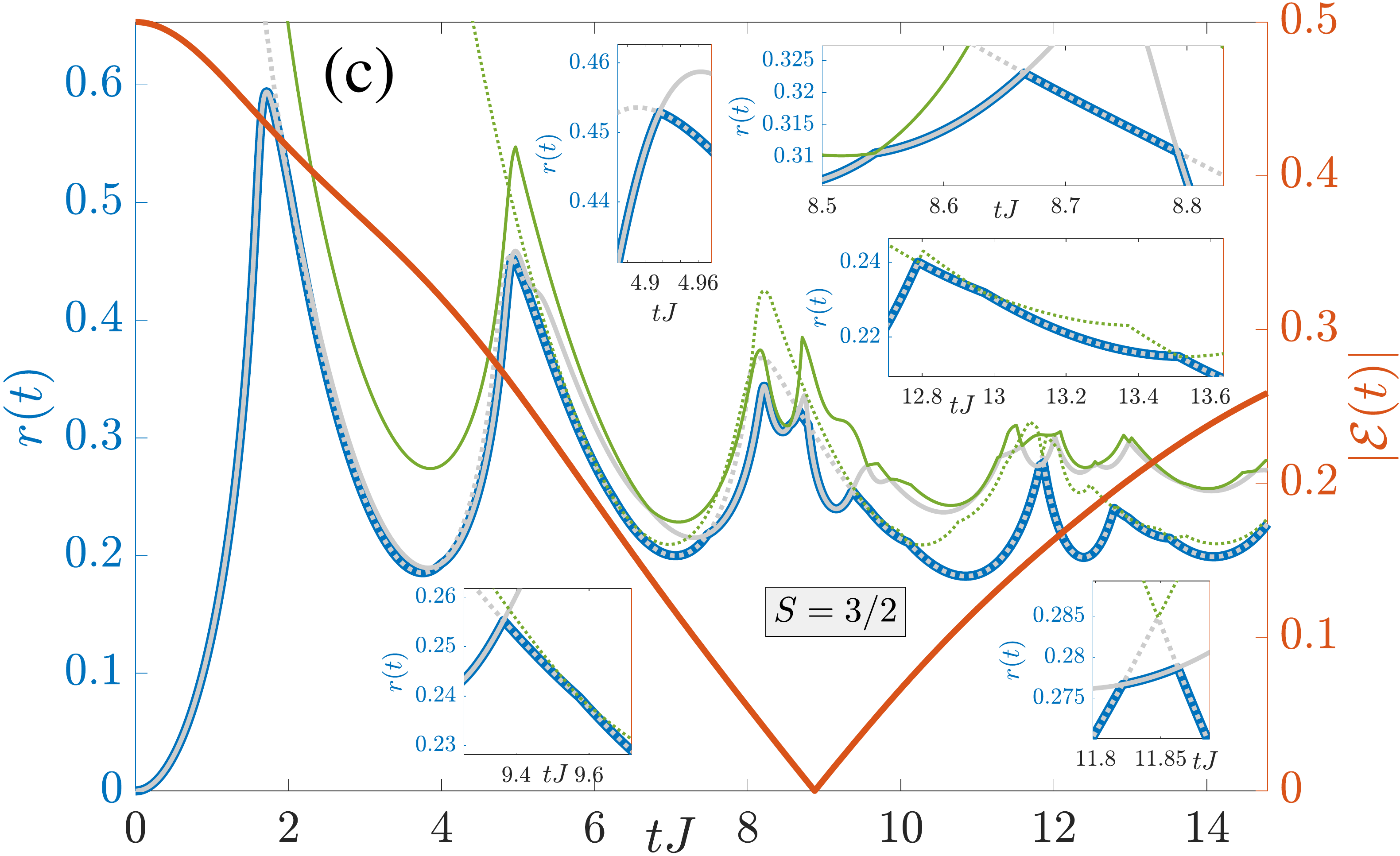}
	\caption{(Color online). Dynamics of the total return rate~\eqref{eq:RR} (blue), which is the minimum of the primary~\eqref{eq:primary_RR} and secondary~\eqref{eq:secondary_RR} return rates (solid and dotted gray curves, respectively) and the order parameter~\eqref{eq:flux} (red) for quenches starting in a $\mathbb{Z}_2$ symmetry-broken product state, which is one of two doubly degenerate ground states of the spin-$S$ $\mathrm{U}(1)$ QLM~\eqref{eq:H} at $\kappa=0.1\sqrt{J}$ and $\mu\to\pm\infty$ (positive sign for half-integer $S$, negative for integer $S$). The solid (dotted) green curves designate the second primary (secondary) rate-function branch; see Sec.~\ref{sec:DQPT_glossary}. (a) For the case of $S=1/2$, periodic manifold DQPTs correspond directly to order-parameter zeros occurring at roughly the same period, in full agreement with the conclusions of Ref.~\cite{Huang2019}. (b,c) For $S>1/2$, this direct connection is no longer present, and many aperiodic manifold and branch DQPTs (see Sec.~\ref{sec:DQPT_glossary} for definition) appear in $r(t)$ even when there is only a single order-parameter zero in the accessible time-evolution window.}
	\label{fig:S_ProdInitState} 
\end{figure}

Let us first focus on the case of $S=1/2$, where we repeat the same quench employed in Ref.~\cite{Huang2019}. The corresponding quench dynamics of the total return rate~\eqref{eq:RR}, its component return rates~\eqref{eq:primary_RR} and~\eqref{eq:secondary_RR}, their second rate-function branches, and the order parameter~\eqref{eq:flux} are shown in Fig.~\ref{fig:S_ProdInitState}(a). In full agreement with Ref.~\cite{Huang2019}, we find within the evolution times accessible in iMPS that the return rate~\eqref{eq:RR}, depicted in blue, exhibits only manifold DQPTs that are directly connected to the order parameter (depicted in red) changing sign.

Interestingly, the primary return rate $\lambda^+_1(t)$~\eqref{eq:primary_RR} (solid gray curve) itself hosts branch DQPTs at half the frequency of the total return rate $r(t)$, where these branch DQPTs are the result of its intersection with $\lambda^+_2(t)$ (solid green line). In other words, if we merely use $\lambda^+_1(t)$ as the total return rate rather than $r(t)$, as is done in many works, then we will also see a direct connection between the emerging (branch rather than manifold) DQPTs and the order-parameter zeros, whereby one such DQPT corresponds to two order-parameter zeros. As such, it does not matter in this case whether the return rate is defined through the dominant branch or manifold, there will always be a direct connection between the resulting DQPTs and the order-parameter zeros.

The question that posits itself here is whether this behavior is special to $S=1/2$ or the particular quench protocol employed. As we will show in the following, we find numerically that this behavior is present only for the case of $S=1/2$ and a quench from a vacuum initial state across the critical point. Let us now employ the same quench protocol but for the case of $S=1$, the corresponding quench dynamics of which are shown in Fig.~\ref{fig:S_ProdInitState}(b). Here we see that the total return rate $r(t)$, displayed in blue, has one manifold DQPT, and three aperiodic secondary branch DQPTs, with only one order-parameter zero during the evolution times we can access in iMPS. The three secondary branch DQPTs in $r(t)$ appear at times $t$ when $r(t)=\lambda^-_1(t)=\lambda^-_2(t)<\lambda^+_1(t)$, where $\lambda^-_1(t)$ is depicted with a dotted gray line, and $\lambda^-_2(t)$ with a dotted green line. This picture is fundamentally different from that of the corresponding case for $S=1/2$, albeit one can still argue that there is a single manifold DQPT appearing in $r(t)$ along with a single zero of the order parameter, and so maybe there is still a direct connection between \textit{manifold} DQPTs and order-parameter zeros.

To check this hypothesis, we consider the same quench protocol for $S=3/2$, with the corresponding dynamics shown in Fig.~\ref{fig:S_ProdInitState}(c). We find that this hypothesis no longer holds, and there is no direct connection between manifold DQPTs and order-parameter zeros. In the accessible time evolution window, we can count at least nine \textit{aperiodic} manifold DQPTs in $r(t)$ while there is only a single order-parameter zero. Moreover, the nonanalytic behavior of $r(t)$ is even richer than in previous cases, with primary branch DQPTs at times $t$ where $r(t)=\lambda^+_1(t)=\lambda^+_2(t)<\lambda^-_1(t)$ and secondary branch DQPTs at times $t$ where $r(t)=\lambda^-_1(t)=\lambda^-_2(t)<\lambda^+_1(t)$.

As such, we have shown that employing the quench protocol of Ref.~\cite{Huang2019} gives rise to a connection between manifold DQPTs and order-parameter zeros only for the case of $S=1/2$, while for $S>1/2$ the dynamical critical behavior is much richer, with a plethora of both manifold and branch DQPTs that show no direct connection to order-parameter zeros.

\subsection{Quench from finite $\mu$ to $-\mu$}\label{sec:mto-m}
\begin{figure}[t!]
	\centering
	\includegraphics[width=.48\textwidth]{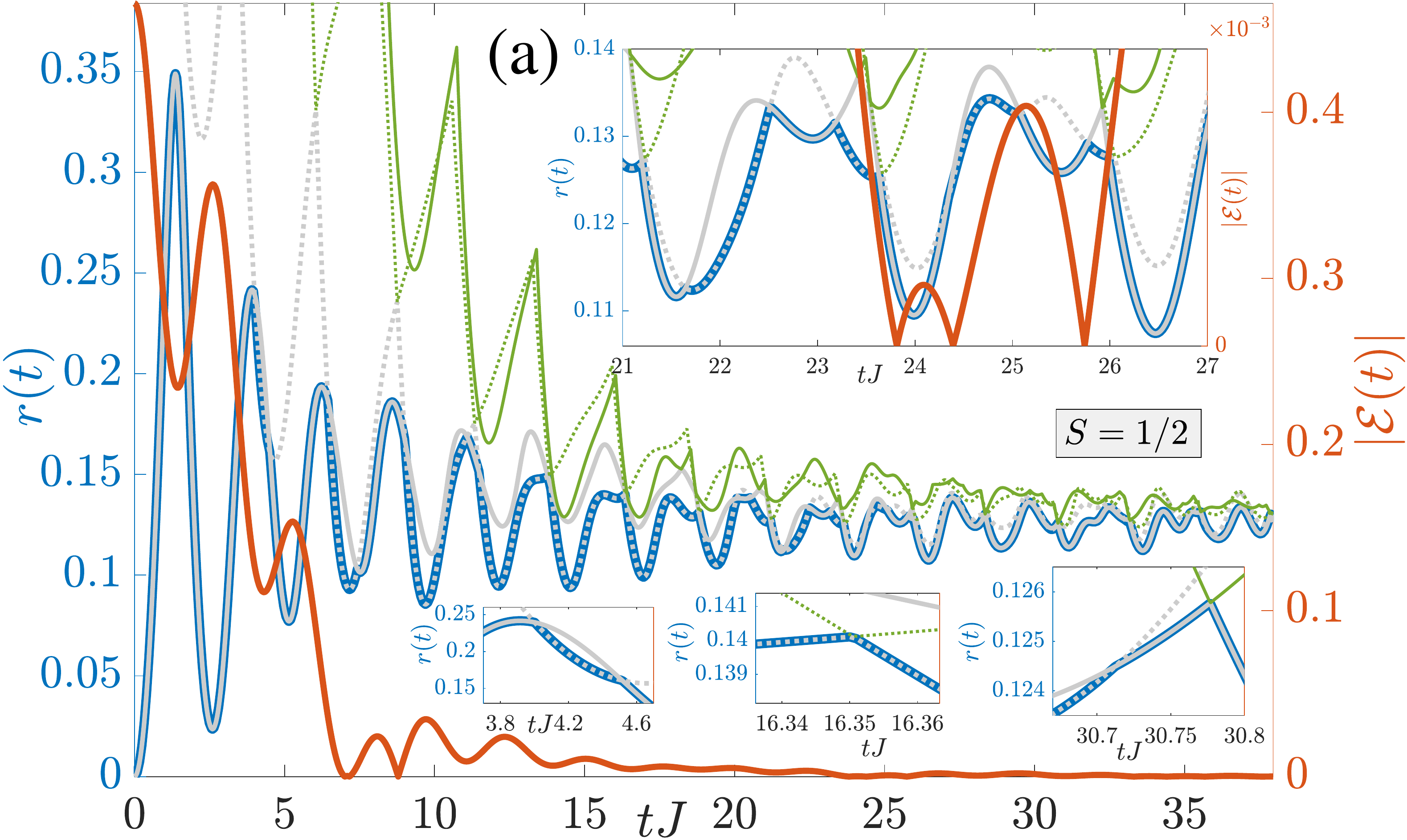}\\
	\vspace{1.1mm}
	\includegraphics[width=.48\textwidth]{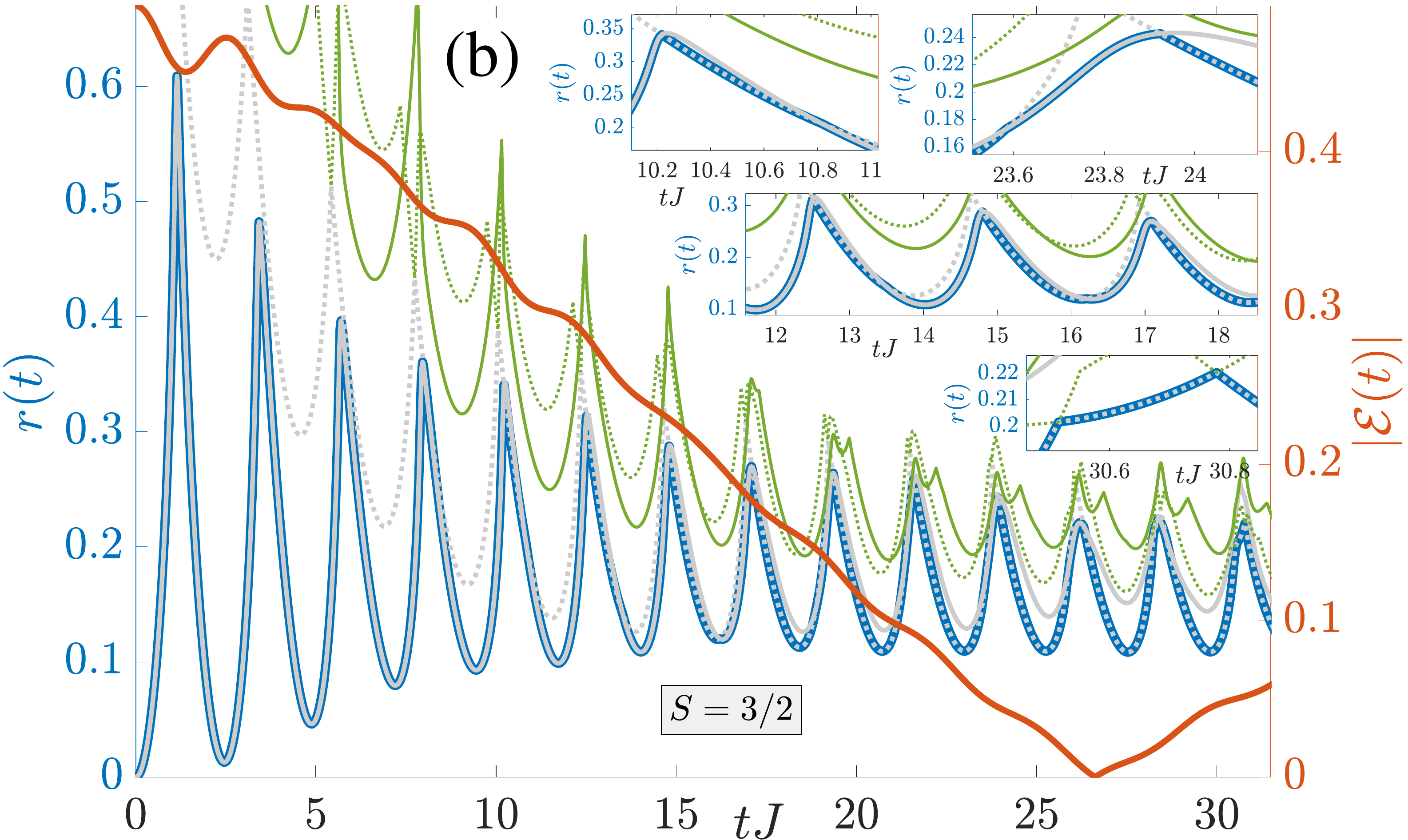}
	\caption{(Color online). Quench dynamics of the return rate~\eqref{eq:RR} and order parameter~\eqref{eq:flux} for the $\mathrm{U}(1)$ QLM with half-integer (a) $S=1/2$ and (b) $S=3/2$ in the wake of a quench from $\mu=J$ to $\mu=-J$ at $\kappa=0.1\sqrt{J}$. As before, this quench is from the $\mathbb{Z}_2$ symmetry-broken phase to the $\mathbb{Z}_2$-symmetric phase of Eq.~\eqref{eq:H} for the case of half-integer $S$. In both cases, there is no direct connection between manifold DQPTs and order-parameter zeros, where we get a plethora of aperiodic manifold DQPTs and several branch DQPTs, but only a few order-parameter zeros within the accessible evolution times.}
	\label{fig:half_integer_S_m1to-1} 
\end{figure}

Let us now consider a different quench protocol, where the initial state is chosen as one of the two degenerate ground states of Eq.~\eqref{eq:H} at $\mu=\pm J$ (positive sign for half-integer $S$, negative for integer $S$), from which the system is quenched with Eq.~\eqref{eq:H} at $\mu=\mp J$ (negative sign for half-integer $S$, positive for integer $S$). The electric-field coupling strength is always set to $\kappa=0.1\sqrt{J}$. The motivation behind this quench is that a sign change in the fermion mass can be interpreted as a change in the topological angle $\theta$ by $\pi$, which has relevance in $(3+1)-$D quantum chromodynamics \cite{Jackiw1976,Peccei1977,Peccei2008,Zache2019,halimeh2021achieving}. Similarly to the quench protocol employed in Sec.~\ref{sec:InitProdState}, the quench here also crosses the equilibrium quantum critical point of the spin-$S$ $\mathrm{U}(1)$ QLM from a $\mathbb{Z}_2$ symmetry-broken phase to a $\mathbb{Z}_2$-symmetric phase for all values of $S$ that we consider in the following.

It is interesting to see if changing the quench protocol from that employed in Sec.~\ref{sec:InitProdState} will alter the picture for $S=1/2$ of a direct connection between manifold DQPTs and order-parameter zeros. As such, we look at the resulting dynamics in the spin-$1/2$ $\mathrm{U}(1)$ QLM in the wake of quenching from $\mu=J$ to $\mu=-J$, shown in Fig.~\ref{fig:half_integer_S_m1to-1}(a). We readily see a breakdown of this picture, with a plethora of manifold DQPTs (at least twenty five) in $r(t)$ occurring over the evolution times we can access in iMPS, during which only six order-parameter zeros appear. In fact, focusing on the evolution times displayed in the largest inset of Fig.~\ref{fig:half_integer_S_m1to-1}(a), we find nine manifold DQPTs in $r(t)$ and only three order-parameter zeros. In addition to manifold DQPTs, we also find over the whole accessible time evolution both primary and secondary branch DQPTs in $r(t)$ (see insets). In contrast to Fig.~\ref{fig:S_ProdInitState}(a), the manifold DQPTs do not occur at a fixed frequency, nor do the order-parameter zeros.

\begin{figure}[t!]
	\centering
	\includegraphics[width=.48\textwidth]{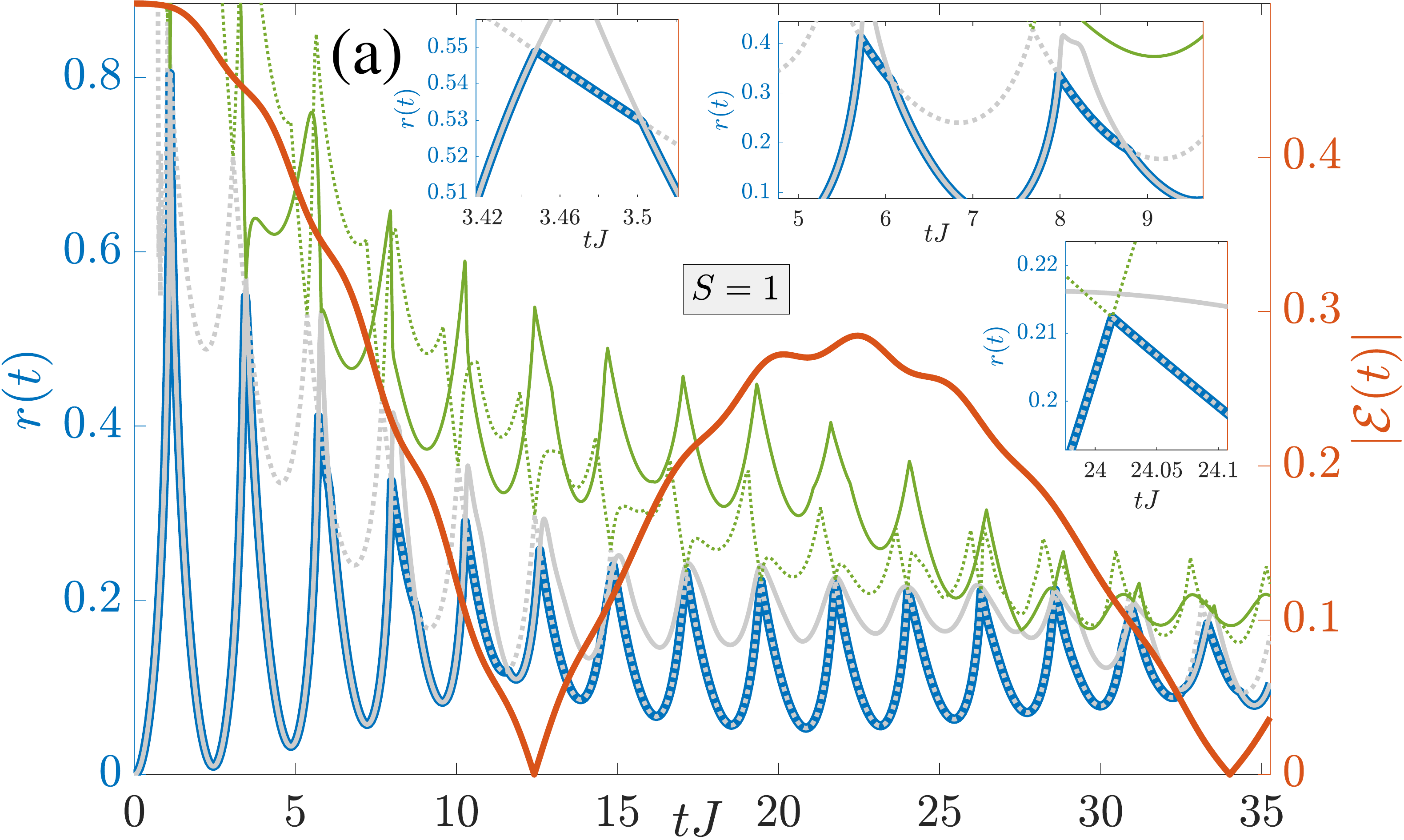}\\
	\vspace{1.1mm}
	\includegraphics[width=.48\textwidth]{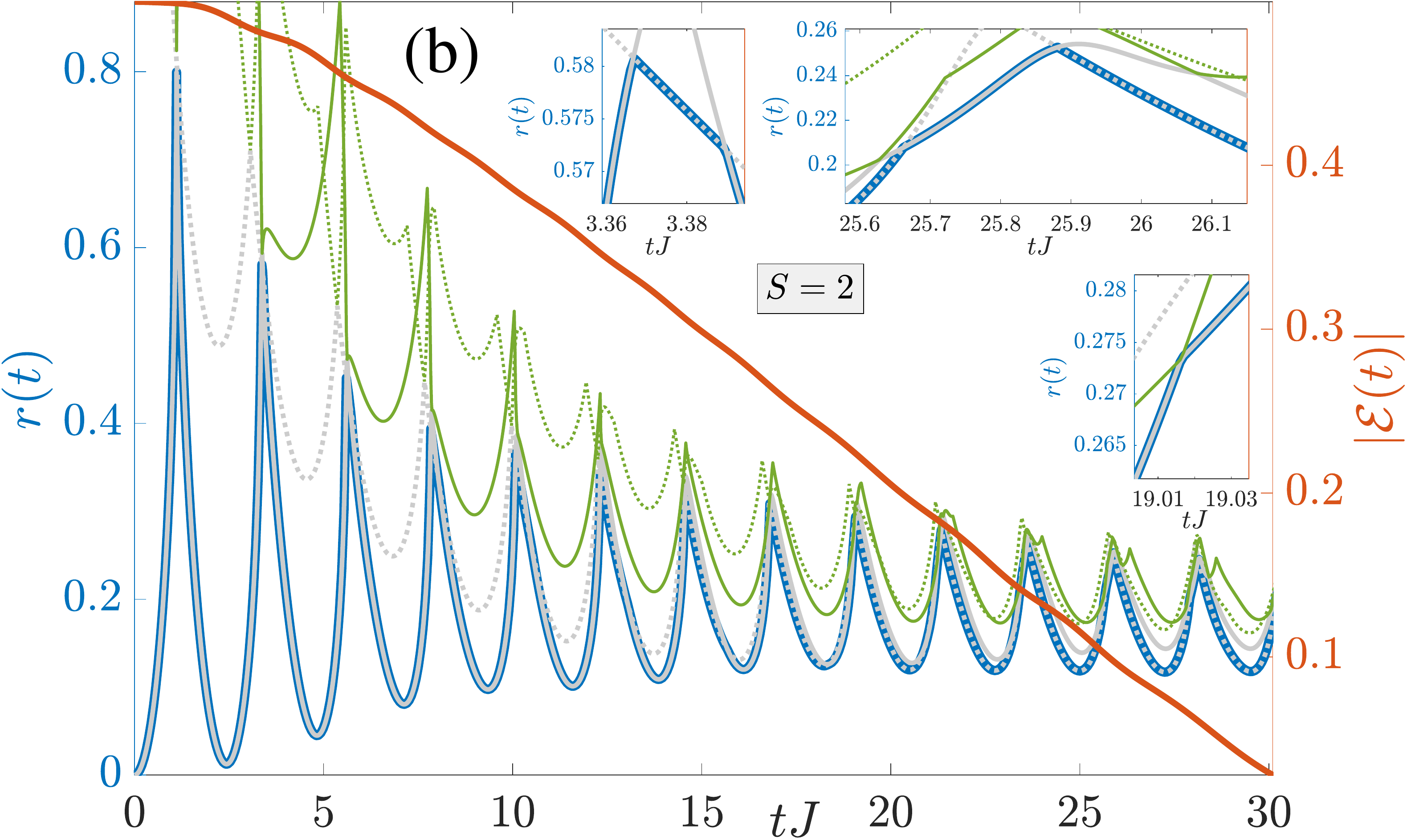}
	\caption{(Color online). Quench dynamics of the return rate~\eqref{eq:RR} and order parameter~\eqref{eq:flux} for the $\mathrm{U}(1)$ QLM with integer (a) $S=1$ and (b) $S=2$ in the wake of a quench from $\mu=-J$ to $\mu=J$ at $\kappa=0.1\sqrt{J}$. As before, this quench is from the $\mathbb{Z}_2$ symmetry-broken phase to the $\mathbb{Z}_2$-symmetric phase of Eq.~\eqref{eq:H} for the case of integer $S$. In both cases, there is no direct connection between manifold DQPTs and order-parameter zeros, where we get a plethora of aperiodic manifold DQPTs and several branch DQPTs, but only a few (or no) order-parameter zeros within the accessible evolution times.}
	\label{fig:integer_S_m-1to1} 
\end{figure}

We consider the same quench but for $S=3/2$ in Fig.~\ref{fig:half_integer_S_m1to-1}(b). Once again, we see a myriad of manifold DQPTs (at least twenty four) in the evolution times accessed in iMPS, along with five secondary branch DQPTs, while the order parameter changes sign only once. The manifold DQPTs also do not occur at a fixed period in time. It is interesting to note that the order-parameter zero occurs at $t\approx26.643/J$, but already at much earlier times (see long inset) we see the return rate $r(t)$ showing somewhat regular manifold DQPTs in its evolution over that early temporal range, without any corresponding zeros in the order parameter during or close to these times. This again indicates that there is no direct connection between DQPTs and order-parameter zeros in general.

Turning to integer $S$, we now consider the quench from $\mu=-J$ to $\mu=J$, which is also from the $\mathbb{Z}_2$ symmetry-broken phase to the $\mathbb{Z}_2$-symmetric phase of Eq.~\eqref{eq:H}. As before, $\kappa=0.1\sqrt{J}$. The corresponding quench dynamics for $S=1$ are displayed in Fig.~\ref{fig:integer_S_m-1to1}(a). In the evolution times accessed by iMPS, we find eighteen aperiodic manifold DQPTs, along with five secondary branch DQPTs, while only two order-parameter zeros exist in the same time interval. On the other hand, we see no zeros of the order parameter in the accessible times for the case pf $S=2$ shown in Fig.~\ref{fig:integer_S_m-1to1}(b), but there are twenty three aperiodic manifold DQPTs and three primary branch DQPTs. Even though we expect that at longer evolution times not accessible in our codes the order parameter may still change sign, our numerical results strongly suggest that there is no direct link between DQPTs and the order-parameter zeros.

It is interesting to note that the dynamics of the spin-$S$ $\mathrm{U}(1)$ QLM after a quench from $\mu$ to $-\mu$ has been shown to converge to the Wilson--Kogut--Susskind (WKS) limit already at small values of the link spin length $S\gtrsim2$ \cite{halimeh2021achieving}. Furthermore, at small values of the electric-field coupling strength $\kappa$ like we consider here, the dynamics is qualitatively the same between half-integer and integer $S$. This is because a small $\kappa$ will not sufficiently suppress quantum fluctuations, thereby not allowing the structure of the spin-$S$ operator to be resolved. As a consequence, the conclusions we obtain in this work for the larger values of $S$ are indicative of DQPT behavior in the WKS limit, suggesting that the direct connection of manifold DQPTs to order-parameter zeros may not extend itself to the quantum field theory limit of the spin-$S$ $\mathrm{U}(1)$ QLM.

Also worth noting is the similarity in our conclusion to DQPT behavior in other many-body systems, such as quantum Ising models. Various works on DQPTs in quantum Ising chains with exponentially decaying interactions \cite{Halimeh2018a} and two-dimensional quantum Ising models \cite{Hashizume2018} show that the direct connection between DQPTs and order-parameter zeros occur only for large quenches from the symmetry-broken phase to somewhere deep in the symmetric phase, where the resulting DQPTs and order-parameter zeros also share the same period. However, for smaller quenches, aperiodic DQPT behavior arises, even when the order parameter itself shows periodic zeros, or none at all \cite{Halimeh2017}.

\section{Conclusion and outlook}\label{sec:conc}
In summary, we have performed numerical simulations of quench dynamics in the spin-$S$ $\mathrm{U}(1)$ quantum link model using the infinite matrix product state technique based on the time-dependent variational principle. We have shown that for generic quenches from the $\mathbb{Z}_2$ symmetry-broken phase to the $\mathbb{Z}_2$-symmetric phase of this model, there is no direct connection between DQPTs and the order-parameter zeros in general, regardless of the value of $S$. Even when starting in product states of the $\mathbb{Z}_2$ symmetry-broken phase and quenching across the equilibrium quantum critical point, only the case of $S=1/2$ shows a direct connection between the occurrence of a DQPT and a sign change in the order parameter in agreement with Ref.~\cite{Huang2019}, but for $S>1/2$, this direct connection is no longer present.

We have shown the existence of two main types of DQPTs in this process. The first is manifold DQPTs, which are the ones occurring at critical times when the total return rate switches between its two component (primary and secondary) return rates onto each of the two degenerate ground states, one of which is the initial state of our system. Manifold DQPTs show a direct connection to order-parameter zeros in the case of $S=1/2$ for a quench from a vacuum initial state across the critical point. The other main type of DQPTs in this work are the branch DQPTs, which are nonanalyticities in the dominant (i.e., lower) component return rate. These are classified into primary branch DQPTs when these nonanalyticities occur in the component (primary) return rate onto the initial state, and secondary branch DQPTs when they occur in the component (secondary) return rate onto the second degenerate ground state. In contrast to their manifold counterparts, branch DQPTs are the direct result of level crossings in the corresponding matrix product state transfer matrix.

Relevant to recent work on the convergence of the spin-$S$ $\mathrm{U}(1)$ quantum link model to the Wilson--Kogut--Susskind limit already at small values of $S\gtrsim2$, we have argued that our results strongly indicate that DQPT behavior in the lattice-QED limit will in general not show a direct connection between DQPTs and order-parameter zeros for quenches from the symmetry-broken phase across the critical point.

Given recent experimental advances in the observation of DQPTs \cite{Jurcevic2017,Flaeschner2018} and quench dynamics of $\mathrm{U}(1)$ quantum link models \cite{Zhou2021}, our work provides a blueprint for future experiments on the dynamical critical behavior of lattice gauge theories.

\begin{acknowledgments}
J.C.H.~acknowledges fruitful discussions with Mari Carmen Bañuls, Haifeng Lang, and Ian P.~McCulloch. This project has received funding from the European Research Council (ERC) under the European Union’s Horizon 2020 research and innovation programm (Grant Agreement no 948141) — ERC Starting Grant SimUcQuam. This work is part of and supported by Provincia Autonoma di Trento, the ERC Starting Grant StrEnQTh (project ID 804305), the Google Research Scholar Award ProGauge, and Q@TN — Quantum Science and Technology in Trento, Research Foundation Flanders (G0E1520N, G0E1820N), and ERC grants QUTE (647905) and ERQUAF (715861). This work was supported by the Simons Collaboration on UltraQuantum Matter, which is a grant from the Simons Foundation (651440, P.Z.).
\end{acknowledgments}

\bibliography{DQPT_U1QLM_biblio}
\end{document}